\documentclass[twocolumn]{aastex701}

\usepackage{amsmath}
\usepackage{amssymb}
\usepackage{xcolor}

\newcommand{\Evec}{\mathbf{E}}
\newcommand{\rvec}{\mathbf{r}}
\newcommand{\Asec}{\mathbf{A}_{\rm sec}}
\newcommand{\Fstoch}{\mathbf{F}_{\rm stoch}}
\newcommand{\FIM}{\mathcal{F}}     

\shorttitle{Seeing Unseen GW Inspirals at the Galactic Center}
\shortauthors{Amaro Seoane et al.}

\begin{document}

\title{Seeing Unseen Gravitational Wave Inspirals at the Galactic Center}

\author[orcid=0000-0003-3993-3249,gname='Pau',sname='Amaro Seoane']{Pau Amaro Seoane}
\affiliation{Universitat Politècnica de València, Spain}
\affiliation{Max Planck Institute f\"ur extraterrestrische Physik, Garching, Germany}
\affiliation{Kavli Institute for Astronomy and Astrophysics, Peking University, Beijing, China}
\email[show]{amaro@upv.es}

\author[orcid=0009-0004-2339-9795,gname='Qimin',sname='Song']{Qimin Song}
\affiliation{Universitat Politècnica de València, Spain}
\affiliation{International Centre for Theoretical Physics Asia-Pacific (ICTP-AP), University of Chinese Academy of Sciences (UCAS), Beijing, China}
\affiliation{Taiji Laboratory for Gravitational Wave Universe (Beijing/Hangzhou), University of Chinese Academy of Sciences (UCAS), Beijing, China}
\email[show]{songqimin23@mails.ucas.ac.cn}

\author[orcid=0000-0003-2357-3787,gname='Shao-Dong',sname='Zhao']{Shao-Dong Zhao}
\affiliation{%
  Institute of Theoretical Physics \& Research Center of Gravitation, 
  School of Physical Science and Technology, Lanzhou University, 
  Lanzhou 730000, China
}
\affiliation{%
  Lanzhou Center for Theoretical Physics, 
  Key Laboratory of Theoretical Physics of Gansu Province, 
  Key Laboratory of Quantum Theory and Applications of MoE, 
  Gansu Provincial Research Center for Basic Disciplines of Quantum Physics, 
  Lanzhou University, Lanzhou 730000, China
}
\email{m15091719613@163.com}

\author[gname='Kostas',sname='Tzanavaris']{Kostas Tzanavaris}
\affiliation{Max Planck Institute for Gravitational Physics (Albert Einstein Institute), Hannover, D-30167, Germany}
\affiliation{Leibniz Universit{\"a}t Hannover, Hannover, D-30167, Germany}
\email{konstantinos.tzanavaris@aei.mpg.de}

\author[gname='Reinhard',sname='Genzel']{Reinhard Genzel}
\affiliation{Max Planck Institute for Extraterrestrial Physics, Giessenbachstra{\ss}e 1, 85748 Garching, Germany}
\affiliation{Departments of Physics \& Astronomy, Le Conte Hall, University of California, Berkeley, CA 94720, USA}
\email{genzel@mpe.mpg.de}

\author[gname='Frank',sname='Eisenhauer']{Frank Eisenhauer}
\affiliation{Max Planck Institute for Extraterrestrial Physics, Giessenbachstra{\ss}e 1, 85748 Garching, Germany}
\affiliation{Department of Physics, TUM School of Natural Sciences, Technical University of Munich, 85748 Garching, Germany}
\email{eisenhau@mpe.mpg.de}

\author[orcid=0000-0003-1572-0396,gname='Thomas',sname='Ott']{Thomas Ott}
\affiliation{Max Planck Institute for Extraterrestrial Physics, Giessenbachstra{\ss}e 1, 85748 Garching, Germany}
\email{ott@mpe.mpg.de}

\author[gname='Stefan',sname='Gillessen']{Stefan Gillessen}
\affiliation{Max Planck Institute for Extraterrestrial Physics, Giessenbachstra{\ss}e 1, 85748 Garching, Germany}
\email{ste@mpe.mpg.de}

\author[gname='Guillaume',sname='Bourdarot']{Guillaume Bourdarot}
\affiliation{Max Planck Institute for Extraterrestrial Physics, Giessenbachstra{\ss}e 1, 85748 Garching, Germany}
\email{gbourda@mpe.mpg.de}

\author[gname='Diogo',sname='Ribeiro']{Diogo C. Ribeiro}
\affiliation{Max Planck Institute for Extraterrestrial Physics, Giessenbachstra{\ss}e 1, 85748 Garching, Germany}
\email{ribeiro@mpe.mpg.de}

\author[gname='Matteo',sname='Sadun-Bordoni']{Matteo Sadun Bordoni}
\affiliation{Max Planck Institute for Extraterrestrial Physics, Giessenbachstra{\ss}e 1, 85748 Garching, Germany}
\email{msadun@mpe.mpg.de}

\author[gname='Simran',sname='Joharle']{Simran Joharle}
\affiliation{Max Planck Institute for Extraterrestrial Physics, Giessenbachstra{\ss}e 1, 85748 Garching, Germany}
\email{sjoharle@mpe.mpg.de}

\author[gname='Felix',sname='Mang']{Felix Mang}
\affiliation{Max Planck Institute for Extraterrestrial Physics, Giessenbachstra{\ss}e 1, 85748 Garching, Germany}
\affiliation{Department of Physics, TUM School of Natural Sciences, Technical University of Munich, 85748 Garching, Germany}
\email{fmang@mpe.mpg.de}

\author[gname='Andreas',sname='Burkert']{Andreas Burkert}
\affiliation{University Observatory, Faculty of Physics, Ludwig-Maximilians-Universit{\"a}t, Scheinerstra{\ss}e 1, 81679 Munich, Germany}
\affiliation{Max Planck Institute for Extraterrestrial Physics, Giessenbachstra{\ss}e 1, 85748 Garching, Germany}
\email{burkert@usm.lmu.de}

\author[gname='Jorge',sname='Cuadra']{Jorge Cuadra}
\affiliation{Universidad Adolfo Ib{\'a}\~nez, Av. Padre Hurtado 750, Vi\~na del Mar, Chile}
\affiliation{Millennium Nucleus on Transversal Research and Technology to Explore Supermassive Black Holes (TITANS), Chile}
\affiliation{Max-Planck-Institut für Astrophysik, Karl-Schwarzschild-Straße 1, 85748 Garching, Germany}
\email{jorge.cuadra@uai.cl}

\author[orcid=0000-0002-9019-9951,gname=Diego, sname='Calderón']{Diego Calderón}
\affiliation{Max-Planck-Institut für Astrophysik, Karl-Schwarzschild-Straße 1, 85748 Garching, Germany}
\email{calderon@MPA-Garching.MPG.DE}

\author[gname="Re'em",sname='Sari']{Re'em Sari}
\affiliation{Racah Institute of Physics, The Hebrew University, Jerusalem 91904, Israel}
\email{sari@mail.huji.ac.il}

\begin{abstract}
The Galactic Center is expected to host a population of extremely large mass-ratio inspirals (X-MRIs) migrating toward the supermassive black hole. We propose an observational approach utilizing the secular gravitational torque exerted by X-MRIs on the highly eccentric S-star S301. This approach is intrinsically limited by the stochastic nature of the surrounding stellar-mass black hole cusp. As S301 propagates through this granular medium, it experiences discrete gravitational encounters. Standard orbital fitting procedures inherently assume a smooth background potential; applying them to stochastic trajectories averages these discrete velocity discontinuities into a continuous secular drift. This introduces a systematic bias that mimics the X-MRI precession signal, invalidating traditional mass estimation techniques. To resolve this degeneracy, we implement a methodology based on the Wavelet Transform Modulus Maxima method to characterize the local {H\"older} regularity of the orbital motion. This model isolates the singular discontinuities induced by stellar-mass black hole encounters from the continuous torque of the X-MRI distribution, allowing for detection even when instantaneous accelerations are below the instrumental noise threshold. We demonstrate analytically that S301 is highly sensitive to the interior X-MRI population while remaining dynamically decoupled from the secular influence of the exterior cusp. Combining precision astrometry and radial velocity projections, {we establish that a minimal population of two X-MRIs is detectable within a five to nine-year timeframe, allowing us to constrain the presence of an unseen orbiter population prior to the launch of space-based gravitational wave detectors.}
\end{abstract}

\keywords { \uat {Galactic center}{565} --- \uat
{Black holes}{162} --- \uat {Stellar dynamics}{1596} ---
\uat {Gravitational waves}{678} --- \uat
{Astrometry}{80} --- \uat {Wavelet
analysis}{1918}}

\section {Introduction} \label{sec:intro}

The center of our galaxy is a dense environment where stars and stellar remnants orbit a supermassive black hole (SMBH), Sgr A*. {High-precision monitoring of these stars has already confirmed predictions of General Relativity, such as the Schwarzschild precession of S2 \citep{GRAVITY2020,GRAVITY2024}. The highly eccentric S-star S301 \citep{S3012026} is particularly well suited as a probe for this environment.} This region is expected to host hidden populations of compact objects. Deep in the gravitational potential, there may be a population of brown dwarfs (BDs) {and very low-mass red dwarfs (M-dwarfs near the hydrogen-burning limit)} inspiralling toward the SMBH, known as extremely large mass-ratio inspirals \citep[X-MRIs,][]{Amaro-Seoane2019,VazquezEtAl2022,VazquezEtAl2023,ColpiEtAl2024,AmaroSeoaneZhao2025,WangEtAl2026,FanEtAl2026,VazquezEtAl2025}. Further out, a cusp of stellar-mass black holes (SBHs) is predicted to exist, some in the early phases of inspiral \citep[early-stage extreme mass-ratio inspirals, or E-EMRIs,][]{AmaroSeoaneEtAl2024mono,AmaroSeoaneZhao2025}.

The collective mass of the X-MRIs, if slightly misaligned, will exert a torque on an S-star. Because the X-MRIs orbit rapidly, their influence is largely averaged over time, resulting in a steady, smooth precession of the S-star's orbit. This is the signal we seek.

The challenge lies in the environment's noise. The surrounding population of SBHs is not smooth; it is granular. As the S-star moves through this population, it receives random gravitational kicks (\citealt{SadunBordoniEtAl2025}). These stochastic thrusts alter the star's trajectory abruptly.

This leads to a fundamental ambiguity. If we observe an S-star's orbit drifting, how do we know if it is the steady pull of the X-MRI population or merely the aftermath of a random kick from a passing SBH? Standard orbital fitting methods assume a smooth path and cannot reliably distinguish between a gentle, continuous pull and a sudden, sharp kick. Attempting to fit a smooth model to a path that has been disrupted by granularity leads to biased results (\citealt{SadunBordoniEtAl2025}).

This paper introduces a system to resolve this ambiguity by analyzing the ``texture" of the orbital motion. A steady pull results in a smooth path, while a kick creates a sharp discontinuity in the velocity. We employ wavelet analysis, which acts as a mathematical microscope capable of zooming in on localized events and characterizing their smoothness, even in the presence of measurement noise. This allows us to identify the signature of a kick and separate it from the signature of a smooth drift.

By integrating this consistent methodology with detailed astrophysical modeling and precision forecasts, we analyze the prospects for detecting the X-MRI population using the star S301. We show that this approach allows for the detection of X-MRIs years before they might be observed using gravitational waves.

\section {System Configuration and Dynamics}
\label{sec:config}

We adopt standard values for the central object's mass $M_{\rm BH} = 4.3\times 10^6\,M_{\odot}$ and distance $D=8.3$ kpc {\citep{GRAVITY2020, GRAVITY2024}}. We analyze distinct populations of unseen perturbers (subscript $p$). As observational probes, we use S2 ($a_{\rm S2} \approx 1030$ AU, $e_{\rm S2} \approx 0.8846$) as a baseline \citep{GRAVITY2020}. {We focus on the star S301 \citep{S3012026}, which has a semi-major axis $a_{1} \approx 687$ AU and an eccentricity $e_{1} \approx 0.9825$.} S301's mean motion is $n_1 \approx {0.724}$ rad/yr (period $T_1 \approx {8.7}$ yr).

The X-MRI population is modelled with a total mass $M_{\rm X}$, distributed as $dN/da_p \propto a_p^{-1/2}$ up to a critical semi-major axis, $a_{\rm crit}$. This radius defines the boundary where dynamics transition from stellar relaxation domination to gravitational radiation domination \citep[following][]{Amaro-SeoaneEtAl07, Amaro-Seoane2019}. We adopt $a_{\rm crit}=600$ AU (period $T_X \approx 7.1$ yr) and assume $e_p=0.95${, which is a characteristic value for objects brought into the inner region \citep{Amaro-Seoane2019}}. When analyzing specific populations $N_X$, we assume a typical brown dwarf mass $m_{\rm BD} \approx 0.062 M_\odot$; {this is a representative mass for X-MRIs, establishing our baseline scalings} \citep{Amaro-Seoane2019}. {Unlike main-sequence stars such as S301---which have low mean densities and are tidally disrupted at pericenters of a few AU long before reaching the gravitational-wave regime---electron-degenerate BDs are sufficiently compact ($R_{\rm BD} \approx 0.08\,R_\odot$) to survive tidal forces right down to the event horizon, locking onto deterministic gravitational-wave inspirals once inside $a_{\rm crit}$. Furthermore, inside $600$ AU at high eccentricities, $10\,M_\odot$ SBHs inspiral via gravitational radiation $\sim 160$ times faster than BDs ($\tau_{\rm GW} \propto m_p^{-1}$) and rapidly merge with Sgr A*, leaving BDs as the dominant long-lived population of discrete, highly eccentric inspirals in this inner cavity \citep{Amaro-Seoane2019}.} {Detailed Fokker--Planck and $N$-body treatments accounting for the initial mass function and mass segregation demonstrate that while heavier remnants sink efficiently, their rapid gravitational-wave merger rates deplete their steady-state numbers inside $a_{\rm crit}$, whereas compact sub-stellar objects and very low-mass M-dwarfs ($\sim 0.08\text{--}0.15\,M_\odot$) maintain a higher steady-state occupation fraction \citep{Amaro-Seoane2019,AmaroSeoaneZhao2025}. Adopting a minimal population of $N_X = 2$ to $5$ BDs with $m_{\rm BD} \approx 0.062\,M_\odot$ sits comfortably below the observational upper limit for extended dark mass within the orbit of S301 \citep{GRAVITY2020,GRAVITY2024} and provides a conservative lower bound on the interior X-MRI signal, as any surviving low-mass M-dwarfs or white dwarfs inside $600$ AU would amplify the secular torque and shorten the required detection timescale.}

Exterior to $a_{\rm crit}$ and S301 lies the stellar cusp, dominated by a granular population of SBHs. This population includes E-EMRIs. We model them as $M_{\rm SBH} = 10 M_\odot$ objects with high eccentricity ($e_p=0.95$){. This represents {typical dynamically relaxed} orbits {near the loss cone} \citep[see e.g.][]{AmaroSeoaneZhao2025}, and importantly, maximizes the secular perturbation on the probe, providing a conservative upper limit for the noise}. This population is the primary source of stochastic gravitational noise (\citealt{SadunBordoniEtAl2025}). We also analyze individual perturbers located at characteristic distances, e.g., $a_p = 1500$ AU {(representing the inner boundary of the massive SBH cusp)}, to assess their secular impact.

The spatial configuration of the perturber orbits determines the net perturbation. While dynamical processes favor isotropy over long timescales \citep{RT96}, an isotropic distribution{---such as a smooth, continuous particle dark matter spike inside $600$ AU---}yields a negligible net quadrupole precession signal{, inducing only an in-plane monopole drift}. We therefore analyze anisotropic configurations, which arise naturally from statistical fluctuations, especially in small {discrete} populations ($N=2$ to 5) {occupying the inner cavity}. {We note that one or a few faint main-sequence stars (such as G, K, or early M-dwarfs) orbiting inside $600$ AU could currently evade near-infrared detection limits while dominating the local quadrupole moment; although such stars have larger radii and would be tidally disrupted before reaching the LISA band (and thus would not become X-MRIs), detecting their secular torque would still place valuable constraints on the unobservable inner stellar distribution.}

\section {Orbital Precession: Theory and Observation} \label{sec:methods}

We employ secular perturbation theory to calculate the smooth, orbit-averaged precession, $\dot{\omega}$. This must be distinguished from relativistic effects, such as the Lense-Thirring (LT) precession. For S301, $\dot{\omega}_{\rm LT, 1} \approx {22.7}\,\chi\, \mathrm{arcsec\,yr}^{-1}${, where $\chi$ is the dimensionless spin parameter of Sgr A*}.

If a perturber population forms an anisotropic distribution, its quadrupole moment induces a secular precession. We model this using the Fisher distribution \citep{Fisher1920,Fisher1922,Fisher1925}, parameterized by the concentration parameter $\kappa$ (related to the RMS inclination spread $\sigma_{I}$).

\subsection{Precession of {an Exterior Probe (S301) induced by Interior Perturbers (X-MRIs)}}

We first consider the probe orbiting exterior to the distribution ($a > a_{\rm crit}$), relevant for S301 perturbed by X-MRIs. The averaged disturbing function $\mathcal{R}$ governs the evolution (e.g., \citealt{MurrayDermott1999}):
\begin{equation}
\mathcal{R}_{\rm ext} = \frac{G m_p}{4 a} \frac{a_p^2}{a^2} (1-e^2)^{-3/2} F_{\rm ext}(e_p) P_2(\cos I_p).
\label{eq:R_ext_single}
\end{equation}
{Here, $G$ is the gravitational constant, $m_p$ is the mass of the perturber, $P_2$ is the second Legendre polynomial, and $I_p$ is the mutual inclination. Note that $e$ is the eccentricity of the observed probe, while $e_p$ is the eccentricity of the unseen perturber.} Furthermore, $F_{\rm ext}(e_p) = 1 + \frac{3}{2}e_p^2$.

Using Lagrange's planetary equations and integrating over the distribution (where $\langle a_p^2 \rangle = a_{\rm crit}^2/5$), the total exterior precession is:
\begin{equation}
\dot{\omega}_{\rm N, ext} = \frac{3n}{20} \left(\frac{M_{\rm X}}{M_{\rm BH}}\right) \frac{a_{\rm crit}^2}{a^2 (1-e^2)^{2}} F_{\rm ext}(e_p) \boldsymbol{\mathcal{R}}_I(\kappa),
\label{eq:quad_precession_ext_disk}
\end{equation}

\noindent
{where $n$ is the mean motion of the probe. The inclination reduction factor $\boldsymbol{\mathcal{R}}_I(\kappa)$ represents the expectation value over the Fisher distribution with concentration parameter $\kappa$.} This equation represents the secular (long-term average) precession. Because the orbital periods of S301 ($T_1 \approx {8.7}$ yr) and the X-MRIs ($T_X \approx 7.1$ yr) are comparable, the true signal will also contain ``short-period terms''---oscillations superimposed on the secular drift, reflecting the specific orbital phases of the few X-MRIs. While we focus on the dominant secular term here, this underlying periodic structure could potentially aid detection.

The reduction factor $\boldsymbol{\mathcal{R}}_I(\kappa) = \langle P_2(\cos I_p) \rangle$ quantifies the effect of misalignment. We calculate this expectation value under the Fisher distribution \citep{Fisher1925}:
\begin{equation}
f(I_p; \kappa) = \frac{\kappa}{2 \sinh \kappa} \sin(I_p) e^{\kappa \cos(I_p)} .
\label{eq:fisher_distribution}
\end{equation}
The reduction factor is the integral:
\begin{equation}
\boldsymbol{\mathcal{R}}_I(\kappa) = \int_0^\pi P_2(\cos I_p) f(I_p; \kappa) dI_p.
\label{eq:R_factor_definition}
\end{equation}
Substituting $P_2(x) = \frac{1}{2}(3x^2 - 1)$ and changing variables to $x = \cos I_p$:
\begin{equation}
\boldsymbol{\mathcal{R}}_I(\kappa) = \frac{\kappa}{4 \sinh \kappa} \int_{-1}^1 (3x^2 - 1) e^{\kappa x} dx.
\end{equation}
This integral is solved exactly through integration by parts, yielding

\begin{equation}
\boldsymbol{\mathcal{R}}_I(\kappa) = 1 - \frac{3\coth\kappa}{\kappa} + \frac{3}{\kappa^2}.
\label{eq:R_factor_exact}
\end{equation}
In the limit $\kappa \gg 1$, $\sigma_{I}^2 \approx 2/\kappa$ (in radians){, where $\sigma_I$ is the root-mean-square inclination spread}.

\subsection{Secular Perturbations {on an Interior Probe by Exterior Populations}: Suppression of the E-EMRI/SBH Signal}
\label{sec:exterior_suppression}

We now analyze the case where the probe (S301) orbits interior to the perturber (SBH/E-EMRI), i.e., $a < a_p$. This configuration leads to a strong suppression of the secular precession signal, which we derive here.

The disturbing function for an interior probe caused by an exterior perturber $m_p$, expanded to quadrupole order and averaged over both orbits (assuming co-planarity for simplicity, $I_p=0$ {---which maximizes the effect, since any mutual inclination would introduce a reduction factor $P_2(\cos I_p) \le 1$, meaning coplanarity provides a robust upper bound for the noise---}), is:
\begin{equation}
\mathcal{R}_{\rm int} = \frac{G m_p}{4 a_p} \frac{a^2}{a_p^2} (1-e_p^2)^{-3/2} F_{\rm int}(e).
\label{eq:R_int_single_math}
\end{equation}
Here, $F_{\rm int}(e) = 1 + \frac{3}{2}e^2$. The crucial difference compared to Eq.~\eqref{eq:R_ext_single} is the dependence on the semi-major axes: $\mathcal{R}_{\rm int} \propto a^2/a_p^3$, whereas $\mathcal{R}_{\rm ext} \propto a_p^2/a^3$.

The precession rate $\dot{\omega}$ is derived using Lagrange's planetary equation for the argument of pericentre:
\begin{equation}
\dot{\omega} = \frac{\sqrt{1-e^2}}{n a^2 e} \frac{\partial \mathcal{R}}{\partial e}.
\label{eq:lagrange_omega}
\end{equation}
We calculate the derivative of $\mathcal{R}_{\rm int}$ with respect to the probe's eccentricity $e$. Since $\partial F_{\rm int}(e)/\partial e = 3e$, we substitute this into Eq.~\eqref{eq:lagrange_omega}:
\begin{equation}
\dot{\omega}_{\rm N, int} = \frac{\sqrt{1-e^2}}{n a^2 e} \left[ \frac{G m_p}{4 a_p^3} a^2 (1-e_p^2)^{-3/2} (3e) \right].
\end{equation}
Simplifying the expression and utilizing Kepler's third law ($n^2 a^3 = G M_{\rm BH}$, assuming $M_{\rm BH} \gg m_p$), we obtain:
\begin{equation}
\dot{\omega}_{\rm N, int} = \frac{3}{4} n \left(\frac{m_p}{M_{\rm BH}}\right) \left(\frac{a}{a_p}\right)^3 \sqrt{1-e^2} (1-e_p^2)^{-3/2}.
\label{eq:quad_precession_int_math}
\end{equation}

The critical insight lies in the scaling with the ratio of semi-major axes: $\dot{\omega}_{\rm N, int} \propto (a/a_p)^3$. Since $a < a_p$, this term leads to a rapid decay of the perturbation as the perturber moves further away. This derivation confirms that S-stars like S301 are powerful probes for the enclosed X-MRI mass, but are largely insensitive to the secular effects of the distant SBH/E-EMRI population.

\section {Detectability and Parameter Constraints}
\label{sec:detectability}

For S2, the Newtonian signal from X-MRIs is intrinsically weak. Assuming a minimal X-MRI population ($N_X=2$, $M_{\rm X} \approx 0.124 M_\odot$) and perfect alignment, Eq.~\eqref{eq:quad_precession_ext_disk} yields only $\dot{\omega}_{\rm N, S2} \approx {0.0059}\,\mathrm{arcsec\,yr}^{-1}$, well below the theoretical baseline precision for S2, $\sigma_{\dot{\omega},\rm S2} \approx 0.09\,\mathrm{arcsec\,yr}^{-1}$ {(see below)}. Therefore, S2 is unsuitable for this experiment. {Similarly, while S55 has a slightly shorter period ($T \approx 13$ yr) than S2, its eccentricity is substantially lower ($e \approx 0.72$), making its sensitivity factor $(1-e^2)^{-2}$ roughly 30 times smaller than that of S301, rendering it much less sensitive.}

We analyze the detection prospects for S301 perturbed by X-MRIs. The Newtonian precession (using normalized parameters $\tilde{M}_X$ (relative to $M_{\rm X}=0.75\,M_{\odot}$), {$\tilde{a}_{\rm crit} = a_{\rm crit} / 600\,\mathrm{AU}$, and $\tilde{F}_{\rm ext}(e_p) = F_{\rm ext}(e_p) / F_{\rm ext}(0.95)$}) is:
\begin{align}
\dot{\omega}_{\rm N, ext, 1}(\kappa) & \approx {5.82}\, \mathrm{arcsec\,yr}^{-1} \times \nonumber \\
& \tilde{M}_X \tilde{a}_{\rm crit}^2 \tilde{F}_{\rm ext}(e_p) \boldsymbol{\mathcal{R}}_I(\kappa).
\label{eq:gen_wN_S301}
\end{align}

This signal coefficient is about 160 times stronger than for S2. We assess the alignment required for this signal to be significant compared to the Lense-Thirring effect ($B > 10\%/\chi$). The condition requires $\boldsymbol{\mathcal{R}}_I(\kappa) > 0.386$.

Using Eq.~\eqref{eq:R_factor_exact}, we find the minimum concentration $\kappa_{\rm min} \approx 3.50$, corresponding to $\sigma_{I} \lesssim 43.3^\circ$. The signal is remarkably reliable to misalignment. The statistical plausibility of achieving this alignment by chance for $N_X=2$ is high ($P \approx 38.6\%$){, obtained by integrating the Fisher distribution over the required solid angle}.

We now quantify the secular perturbation on S301 from the exterior SBH population using the derivation in Section~\ref{sec:exterior_suppression}. We consider a single $10 M_\odot$ SBH, assuming co-planarity for maximum signal ($P_2=1$).

For a SBH located at $a_p = 1500$ AU ($e_p=0.95$), the suppression factor is $(a/a_p)^3 \approx 0.096$. {Furthermore, as seen in Eq.~\eqref{eq:quad_precession_int_math}, the high-eccentricity ($e_p=0.95$) enhances the precession via the factor $(1-e_p^2)^{-3/2}$. Relaxing to lower eccentricities would reduce this term toward unity, suppressing the effect further and solidifying $e_p=0.95$ as a conservative worst-case scenario.} The induced precession is:
\begin{equation}
\dot{\omega}_{\rm N, int, 1} \approx {0.153}\,\mathrm{arcsec\,yr}^{-1}.
\end{equation}
The secular perturbation from individual exterior SBHs on S301 is weak.

We analyze the achievable precision $\sigma_{\dot{\omega}}$ for S301 by combining astrometry and radial velocity (RV) measurements using the Fisher Information Matrix ($\FIM$). The FIM is a mathematical tool that quantifies how much information the observed data carry about the unknown parameters of a model ($\boldsymbol{\theta}$, the orbital elements). It is defined by the sensitivity of the model predictions ($\mu_i$) to changes in the parameters, weighted by the measurement uncertainties ($\sigma_i$):
\begin{equation}
\FIM_{jk} = \sum_i \frac{1}{\sigma_i^2} \left(\frac{\partial \mu_i}{\partial \theta_j} \right) \left(\frac{\partial \mu_i}{\partial \theta_k} \right).
\label{eq:FIM_definition}
\end{equation}
Physically, the FIM measures the curvature of the likelihood function near its maximum. A large FIM value corresponds to a sharp peak, indicating that the parameters are well-constrained by the data.

The Cramer-Rao bound states that the inverse of the FIM provides a lower bound on the covariance matrix of the parameters: $\mathrm{Cov}(\boldsymbol{\theta}) \ge \FIM^{-1}$. The uncertainty on a specific parameter $\theta_j$ is therefore bounded by:
\begin{equation}
\sigma_{\theta_j} \ge \sqrt{(\FIM^{-1})_{jj}}.
\label{eq:cramer_rao}
\end{equation}
This provides a forecast of the theoretical best-case precision.

The precision scaling laws emerge from the structure of the partial derivatives in Eq.~\eqref{eq:FIM_definition}. We focus on the argument of pericentre, $\omega$. A change in $\omega$ corresponds to a rotation of the Keplerian ellipse within its plane.

For astrometry, the precision depends on how effectively this rotation can be measured against the positional uncertainty $\sigma_{\rm pos}$. The displacement $\Delta r$ caused by a change $\Delta \omega$ is nearly proportional to the semi-major axis $a$. Intuitively, $\Delta r \approx a \Delta \omega$. The precision $\sigma_{\omega}$ is achieved when this displacement is comparable to the measurement uncertainty: $a \sigma_{\omega} \approx \sigma_{\rm pos}$, leading to the scaling $\sigma_{\omega, \rm Astro} \propto \sigma_{\rm pos}/a$.

The eccentricity dependence arises from the geometry of the ellipse. The orientation of the major axis is easier to determine for eccentric orbits (factor of $1/e$), but the width of the ellipse (the semi-minor axis $b = a\sqrt{1-e^2}$) shrinks as $e\to 1$, reducing the leverage perpendicular to the major axis (factor of $\sqrt{1-e^2}$). Combining these factors yields the scaling law
\begin{equation}
\sigma_{\omega, \rm Astro} \propto \left(\frac{\sigma_{\rm pos}}{a}\right) \frac{\sqrt{1-e^2}}{e}.
\label{eq:fisher_scaling_astro}
\end{equation}

For radial velocities, the precision depends on the RV uncertainty ($\sigma_{\rm RV}$) relative to the RV semi-amplitude $K$. The scaling law shares the same geometric eccentricity dependence:
\begin{equation}
\sigma_{\omega, \rm RV} \propto \left(\frac{\sigma_{\rm RV}}{K}\right) \frac{\sqrt{1-e^2}}{e}.
\label{eq:fisher_scaling_rv_K}
\end{equation}
The semi-amplitude scales as $K \propto a^{-1/2} (1-e^2)^{-1/2}$. Substituting this into Eq.~\eqref{eq:fisher_scaling_rv_K}:
\begin{equation}
\sigma_{\omega, \rm RV} \propto \,  \sigma_{\rm RV} \, a^{1/2} \, \frac{(1-e^2)}{e}.
\label{eq:fisher_scaling_rv}
\end{equation}
This scaling reveals a fundamental difference: RV precision improves dramatically for tighter orbits and higher eccentricities, because the high velocities near pericentre provide exceptional leverage to constrain the orientation $\omega$.

We define the geometric improvement factor, $F_{\rm geo}$, relative to S2. $F_{\rm geo, Astro} \approx 1.90$. $F_{\rm geo, RV} \approx 8.58$.

Since the measurements are independent, the total information is the sum: $\FIM_{\rm Total} = \FIM_{\rm Astro} + \FIM_{\rm RV}$. The combined uncertainty $\sigma_{\rm Combined}$ is determined by {the summation} of the inverse variances:
\begin{equation}
\frac{1}{\sigma_{\rm Combined}^2} = \frac{1}{\sigma_{\rm Astro}^2} + \frac{1}{\sigma_{\rm RV}^2}.
\label{eq:inverse_variance_sum}
\end{equation}

We adopt a baseline calibrated to current instrumental capabilities: $\sigma_{\rm pos} = {50\,\mu\mathrm{as}}$ and $\sigma_{\rm RV} = 12$ km/s \citep[VLT spectroscopy,][]{GRAVITY2020}.

We calculate the theoretical baseline precision using the FIM formalism and scale it to S301 over a projected 15-year observation window (using $\sigma_{\dot{\omega}}(T) \propto T^{-3/2}$).
\begin{equation}
\sigma_{\dot{\omega}, \rm S301, Combined} \approx 0.0406\,\mathrm{arcsec\,yr}^{-1}.
\end{equation}
The error is dominated by the RV precision. Following the same procedure for S2 yields $\sigma_{\dot{\omega}, \text{S2}, \text{Combined}} \approx 0.09 \,\mathrm{arcsec\,yr}^{-1}$.

The secular SBH signal ($\approx {0.153}$ arcsec/yr) is potentially detectable (SNR $\approx 3.8$ over 15 years). While weak, it demonstrates that the secular effects of exterior SBHs/E-EMRIs are measurable, though significantly less prominent than the X-MRI signal.

We now analyze the detection timescale for X-MRIs ($T_{4\sigma}$), assuming the strong alignment scenario ($\boldsymbol{\mathcal{R}}_I = 0.386$) and $m_{\rm BD} \approx 0.062 M_\odot$. The timescale scales as $T_{\rm det} \propto (M_{\rm X})^{-2/3}$. {This is obtained by equating the linearly accumulating signal ($\Delta \omega = \dot{\omega}_{\rm N} T$) to the $4\sigma$ detection threshold, where the precision itself scales as $\sigma_{\dot{\omega}} \propto T^{-3/2}$.}

For $N_X=2$ ($M_{\rm X} \approx 0.124 M_\odot$), the expected signal is $\dot{\omega}_{\rm N}(2) \approx {0.371}\,\mathrm{arcsec\,yr}^{-1}$. The required timescale is:
\begin{equation}
T_{4\sigma}(2) \approx 8.7\,\mathrm{years}.
\end{equation}

For $N_X=3$, $T_{4\sigma}(3) \approx 6.6$ years. For $N_X=5$, the timescale reduces substantially to $T_{4\sigma}(5) \approx 4.7$ years. Hence, detecting the minimal X-MRI population is feasible in well under a decade using existing instrumentation.

The Fisher Matrix analysis quantifies the achievable statistical precision ($\sigma_{\dot{\omega}}$), assuming the underlying model (smooth precession) is correct. However, the astrophysical environment is stochastic because of the granularity of the SBH population. The fundamental challenge is ensuring accuracy. A random gravitational thrust from a close encounter with a SBH can mimic a secular drift.

We must assess if these thrusts are detectable above the measurement noise. Using the impulse approximation, $\Delta v \approx 2Gm/(bv_{\rm rel})$. For a $10 M_\odot$ SBH and a typical relative velocity $v_{\rm rel} \approx 500$ km/s, an encounter at an impact parameter $b \approx 270$ AU {(values representative of the typical virial velocity and closest expected encounters in the local SBH density; \citealt{SadunBordoniEtAl2025})} yields a velocity kick $\Delta v \approx 131$ m/s. Over a short interval (e.g., 0.1 yr), this kick induces an astrometric deviation $\Delta r \approx \Delta v \cdot t \approx {0.3\,\mu\mathrm{as}}$, completely buried beneath the single-epoch GRAVITY precision. Nonetheless, over the projected 15-year observation window, this unmodeled velocity shift accumulates into an astrometric deviation $\Delta r \approx {50\,\mu\mathrm{as}}$, equal to the GRAVITY precision limit. Encounters at this distance are frequent enough to significantly impact the orbit (\citealt{SadunBordoniEtAl2025}). Therefore, while instantaneous kicks are hidden by noise, their cumulative effects are observable and perfectly masquerade as smooth secular drift, necessitating a method like WTMM {(Wavelet Transform Modulus Maxima, discussed in Section \ref{sec:wavelets})} to detect them.

Recent detailed studies (\citealt{SadunBordoniEtAl2025}) show that fitting the orbit with a smooth extended potential model, when the true background is granular, leads to highly biased estimates of the enclosed mass. The bias arises when the cumulative noise is dominated by a few discrete events (moderate thrusts) rather than Gaussian fluctuations. These dominant thrusts create localized discontinuities (singularities) in the trajectory. Standard orbital fitting and the FIM analysis rely on assumptions violated by these singularities. If the signal is contaminated, the estimated precession rate will be biased.

Therefore, validating the assumption of smoothness is primary. The wavelet-based framework presented in Section~\ref{sec:wavelets} provides the necessary tools to identify and isolate these dominant stochastic events, even when their amplitude is near the instrumental noise floor, bridging the gap between precision and reliable detection.

\section {Methodology: A Wavelet Framework for Unbiased Measurement}
\label{sec:wavelets}

To address the fundamental ambiguity between the smooth X-MRI signal and the stochastic noise from SBH granularity, we present a structure based on characterizing the local regularity of the orbital signals, robustly accounting for observational noise.

\subsection {The Stochastic Nature of the Signal}

The evolution of the {observed} orbital elements, $\Evec(t)$, is governed by a stochastic differential equation (SDE):
\begin{equation}
\frac{d\Evec}{dt} = \Asec(\Evec, \chi) {+ \mathbf{A}_{\rm S-stars}(\Evec)} + \Fstoch(t) + \mathbf{N}_{\rm obs}(t).
\label{eq:SDE}
\end{equation}
$\Asec$ represents the smooth secular drift. This term arises from interactions averaged over orbital periods (secular theory), such as the X-MRI signal. Even for a few X-MRIs, while short-period oscillations exist, the resulting effect is structurally a continuous torque and smooth evolution. {The term $\mathbf{A}_{\rm S-stars}(\Evec)$ represents the deterministic secular torque from other known S-stars, which is explicitly modeled and subtracted.}

$\Fstoch(t)$ is the stochastic component resulting from SBH granularity. These interactions are impulsive because the encounter timescale ($\tau_{\rm enc} \sim$ days) is much shorter than the orbital period ($T_1 \approx {8.7}$ years). They are modeled as discrete, sharp thrusts.

$\mathbf{N}_{\rm obs}(t)$ is the observational (measurement) noise. A strong thrust causes a discontinuity in the stellar velocity $\dot{\rvec}(t)$.

\subsection {The Wavelet Transform, Local Regularity, and Noise}

We utilize the Continuous Wavelet Transform (CWT) to analyze the orbital residuals $\delta\rvec(t)$, {defined as the spatial deviation of the observed trajectory from a smooth reference orbit} \citep[see e.g.][]{Meyer_1993,daubechies1992ten,chui1992introduction,akansu2001multiresolution}:
\begin{equation}
W(s, \tau) = \int_{-\infty}^{\infty} \delta\rvec(t) \frac{1}{\sqrt{|s|}} \psi^*\left(\frac{t-\tau}{s}\right) dt.
\label{eq:CWT}
\end{equation}
Here, $s$ is the scale {parameter {governing} the wavelet's width, $\psi^*$ represents the complex conjugate of the mother wavelet function (a localized, zero-mean oscillating wave used as a filter),} and $\tau$ is the time localization. {To define the regularity measured by this transform, consider one component $f=f(t)$ of the position residual $\delta\mathbf{r}$. We say that $f$ has a \textit{pointwise H\"older regularity} of order $\alpha>0$ at $\tau_0$ if there is a constant $C>0$ and a polynomial $P_{\tau_0}$ of degree $\leq \alpha$ such that
\begin{equation}
    |f(\tau_0 + \Delta\tau)-P_{\tau_0}(\Delta\tau)|\leq C|\Delta\tau|^\alpha{,}
\end{equation}
for all $\Delta\tau$ sufficiently close to 0. The \textit{H\"older exponent} $h(\tau_0)$ is the supremum of all orders $\alpha$ for which the above bound holds \citep{MallatHwang1992}, and {provides a measure of the signal's local differentiability} at $\tau_0$.}

The diagnostic power stems from the relationship between the local regularity of the signal, quantified by the H\"older exponent $h$, and the asymptotic decay of the wavelet coefficients as $s \to 0$. {{Specifically,} higher values ($h > 1$) correspond to a smooth continuous curve (such as the secular X-MRI torque), whereas lower values ($h \le 1$) indicate sharp, non-differentiable features, such as a velocity discontinuity induced by a stochastic kick. The exponent $h$ characterizes the local differentiability of the signal in the integral of Eq.~\eqref{eq:CWT}, which dictates the asymptotic {decay} of the wavelet coefficients as a function of the scale $s$, effectively measuring the smoothness of the underlying functions.}

The connection arises because wavelets with $N$ vanishing moments act as localized differential operators. The CWT measures the deviation from the local Taylor expansion \citep{Taylor1715}. If $N>h$, the transform is dominated by the singular part of the function.

A fundamental theorem (\citealt{mallat1999wavelet}) formalizes this relationship:
\begin{equation}
|W(s, \tau_0)| \sim s^{h(\tau_0) + 1/2}.
\label{eq:wavelet_decay}
\end{equation}

We must consider the regularity of the different signal components: \begin{enumerate} \item {Smooth Secular Drift ($\Asec${, including S-stars}):} The X-MRI signal is continuous and differentiable ($h>1$). Its coefficients decay rapidly (e.g., $s^{5/2}$ or faster). \item {Gravitational Thrusts ($\Fstoch$):} A thrust causes a discontinuity in velocity ($h=1$). The coefficients decay as $s^{3/2}$. \item {Observational Noise ($\mathbf{N}_{\rm obs}$):} Measurement noise (e.g., White Gaussian Noise) is highly irregular ($h=-1/2$). Its coefficients scale as $s^{0}$. \end{enumerate}

Moreover, because the measurement noise ($h=-1/2$) is less regular than the thrust ($h=1$), the noise will dominate the wavelet spectrum at the smallest scales ($s \to 0$). This prevents a direct measurement of $h$ by simply examining the asymptotic slope across all scales.

\subsection {Detection with Wavelet Transform Modulus Maxima (WTMM)}

To identify singularities in the presence of highly irregular noise, we employ the Wavelet Transform Modulus Maxima (WTMM) method (\citealt{MallatHwang1992}). A singularity at time $t_k$ creates a line of local maxima in the magnitude of the wavelet coefficients, $|W(s, t_k)|$, extending across scales and converging to $t_k$ as $s \to 0$. This is the spine of the ``cone of influence''.

The key insight of the WTMM method is that the decay of the coefficients along these specific maxima lines still follows the power law related to the underlying singularity~{(Eq.~\eqref{eq:wavelet_decay})}, over the range of scales where the signal is detectable above the noise.

This works even if the amplitude of the kick is below the noise level in the raw time domain. The wavelet transform integrates the signal over the wavelet's support (scale $s$). This integration smooths the incoherent noise (providing an integration gain) while enhancing the localized, coherent structure of the singularity.

The WTMM method distinguishes true singularities by identifying coherent maxima lines across multiple scales. While noise also creates local maxima, these do not align coherently.

The analysis proceeds by: (1) Computing the CWT; (2) Identifying and chaining the local maxima across scales to form maxima lines; (3) Analyzing the decay of the coefficients along these lines.

If a coherent line is identified at $t_k$, and the slope of $\log|W(s, t_k)|$ versus $\log(s)$ along this line (at intermediate scales where the signal exceeds the noise floor) is measured to be $3/2$, it unambiguously identifies a velocity discontinuity (a thrust), differentiating it from both the smooth X-MRI signal ($h>1$) and measurement noise ($h=-1/2$).

The first stage is signal classification by the WTMM method. This step characterizes the impact of granularity and handles measurement noise. If a singularity (thrust) is detected at $t_k$, the assumption of global smoothness required by the Fisher Matrix analysis is invalid, preventing the biased results identified by \citet{SadunBordoniEtAl2025}.

The second stage is modelfitting. If contaminated by a thrust, the analyst segments the data at $t_k$ and performs a piecewise fit. This isolates the stochastic event and allows for the recovery of the true underlying secular precession, decoupling the estimation of the X-MRI signal from the discontinuity $\Delta\Evec_k$. This methodology is crucial for ensuring the accuracy of the detection amidst the stochastic background.

\section {Reverse Engineering and Precursor Science for LISA}

A confident detection of anomalous precession, validated by the wavelet framework, provides a pathway to characterize the X-MRI population prior to the LISA mission.

{The total observed precession is:}
\begin{align}
\dot{\omega}_{\rm obs}(a, e) =& \, \dot{\omega}_{\rm LT}(\chi) + \dot{\omega}_{\rm N, ext}(M_{\rm X}) +\nonumber \\
 & \dot{\omega}_{\rm N, int}(M_{\rm SBH}) {+ \dot{\omega}_{\rm S-stars}}.
\label{eq:total_precession_residual}
\end{align}

\noindent
{The term $\dot{\omega}_{\rm S-stars}$ explicitly accounts for the secular effect of other known S-stars like S2 or S55. Because these are bright objects with precisely determined 3D orbits and masses, their secular influence is fully deterministic and can be explicitly forward-modeled and subtracted from the observations to isolate the residual anomalous precession $\Delta\dot{\omega}_{\rm obs}$.} {Because almost all known luminous S-stars orbit exterior to S301 ($a_{\rm S\text{-star}} > 687$ AU), their secular quadrupolar torque on S301 is strongly suppressed by the factor $(a/a_p)^3 \sqrt{1-e^2} \ll 1$ in Eq.~\eqref{eq:quad_precession_int_math} (yielding $\lesssim 0.1\,\mathrm{arcsec\,yr}^{-1}$ even for S2). Consequently, current spectroscopic mass uncertainties ($\sim 10\text{--}20\%$) propagate into a residual subtraction error of $\lesssim 0.01\text{--}0.02\,\mathrm{arcsec\,yr}^{-1}$, well below the X-MRI signal.} As shown previously and derived in Section~\ref{sec:exterior_suppression}, the secular contribution from individual exterior SBHs ($\dot{\omega}_{\rm N, int}$) is negligible for S301 compared to the X-MRI signal.

{While a single measurement of S301's precession rate cannot uniquely break the degeneracies between the number of X-MRIs, their individual masses, and their inclinations{---nor distinguish whether the interior quadrupole is driven purely by compact X-MRIs (BDs and very low-mass M-dwarfs) or by a few faint, undetected main-sequence stars (such as G, K, or early M-dwarfs) inside $600$ AU that would tidally disrupt prior to entering the LISA band---}it serves to robustly prove the existence of an enclosed orbiter population and establish its bulk quadrupolar moment. {Combining these dynamical constraints with deeper near-infrared imaging from upcoming facilities such as GRAVITY+ and MICADO at the ELT will help identify or rule out main-sequence interlopers.} Measuring the exact population properties would require simultaneous multi-probe observations (using multiple S-stars) or tracking the short-period orbital fluctuations.} The degeneracy between $\chi$ and $M_{\rm X}$ can be broken by observing multiple S-stars, because of the distinct functional dependencies on $(a, e)$. $\dot{\omega}_{\rm LT} \propto a^{-3}$ while $\dot{\omega}_{\rm N, ext} \propto a^{-3.5}$.

The connection to {LISA \citep{Amaro-SeoaneEtAl2017,ColpiEtAl2024}, TianQin \citep{TianQin2025,LuoEtAl2016}, and Taiji \citep{Taiji2020}} is significant. The detection by GRAVITY+ and associated spectrographs would provide precursor confirmation of the X-MRI population \citep{Amaro-Seoane2019} years before these observatories begin operations. The constraints on $M_{\rm X}$ (and thus $N_X$) and $a_{\rm crit}$ would validate astrophysical models and provide highly beneficial priors for subsequent data analysis.

\section {Numerical simulation for the precession of S301 induced by {interior} perturbers (X-MRIs)}

\begin{figure}[t]
  \centering
  \includegraphics[width=\columnwidth]{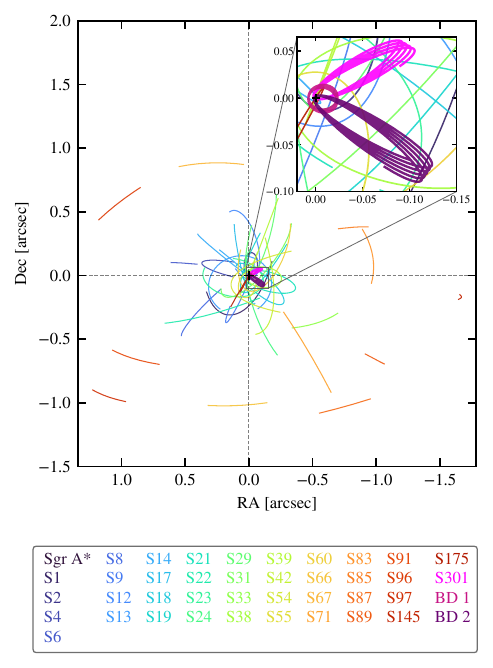}
  \caption{Trajectories of all celestial bodies, including Sgr~A*, 38 S-stars and 2 BDs. The coordinate origin is defined at the initial position of Sgr~A*. The inset provides a zoomed-in view of the trajectories of S301 and the two BDs.}
  \label{fig:trajectories}
\end{figure}

We performed direct $N$-body simulations using ARCHAIN \citep{1999CeMDA..74..287M, 1999MNRAS.310..745M, 2002CeMDA..84..343M, 2008AJ....135.2398M}, including post-Newtonian corrections up to 3 PN, with 2.5 PN gravitational-radiation reaction. We modeled S301 and {37 other S-stars}{, representing the subset of S-stars with the most robust and complete 3D orbital solutions derived from long-term astrometric monitoring,} orbiting the SMBH in {the} GC for about 35 years starting from 1994, as one simulation. {The second simulation is identical, except that it includes} two BDs (X-MRIs), whose distribution follows the prescription presented in \citep{Amaro-Seoane2019,AmaroSeoaneZhao2025}. We depict the 37 S-stars, S301, 2 BDs, and the SMBH at the Galactic Center, with the initial SMBH position chosen as the origin (see Fig.~\ref{fig:trajectories}).

\subsection {Perfect Alignment Case: A Qualitative Analysis of the S301 Precession Rate}

\begin{figure}[t]
  \centering
  \includegraphics[width=\columnwidth]{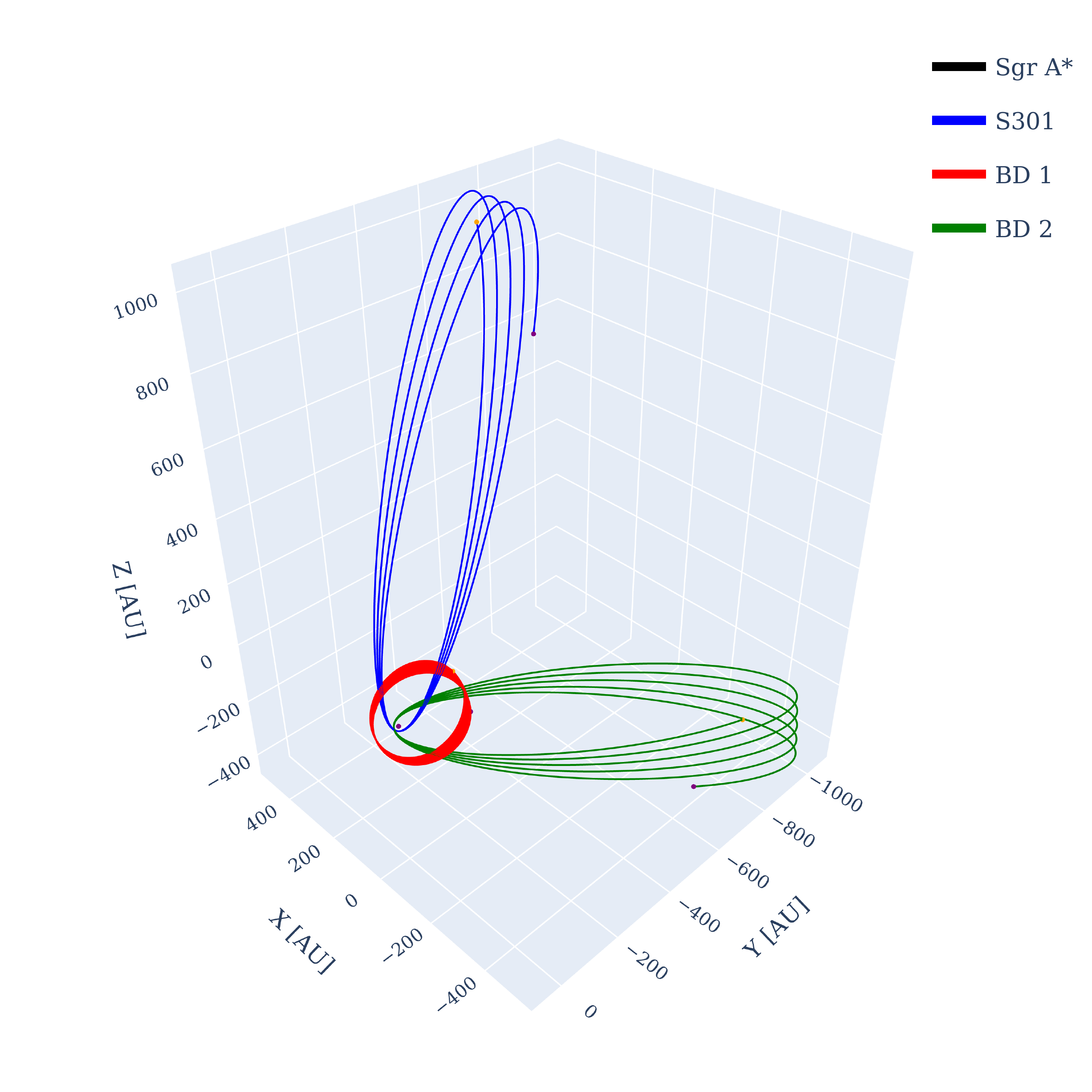}
  \caption{Evolutionary trajectories of the system comprising Sgr~A*, S301, and two BDs. All components are {initialized as} coplanar {(apparent out-of-plane separation is due to the 3D perspective projection and minor out-of-plane perturbations from the 37 background S-stars over 35 years)}.}
  \label{fig:stellar3d}
\end{figure}

We select and show Sgr A*, S301 and two BDs in Fig.~\ref{fig:stellar3d}, as the motions of these objects are of primary interest to us. The semi-major axes and eccentricities of two BDs are $a_1 = 128.57 \mathrm{AU}, a_2 = 600 \mathrm{AU}, e_1 = 0.50, e_2 = 0.98$. Moreover, we consider the orbits of two BDs and S301 {to be initialized} in the same plane, corresponding to perfect alignment, $\boldsymbol{\mathcal{R}}_I = 1$. In this figure, we use orange and purple dots to represent starting and ending points respectively.

We measure the pericentre positions of S301 in simulations with and without X-MRIs, and compare them. Considering an evolutionary timescale of roughly 35 years, S301 completes slightly more than four orbital cycles, making it possible to identify four pericentres.

In order to quantitatively describe the positions of pericentres, we introduce the eccentricity vector \citep{Cordani2003},
\begin{equation}
  \mathbf{e} = \frac{\mathbf{v} \times \mathbf{h}}{\mu} - \frac{\mathbf{r}}{|\mathbf{r}|} = \left(\frac{|\mathbf{v}|^2}{\mu} - \frac{1}{|\mathbf{r}|}\right)\mathbf{r} - \frac{\mathbf{r} \cdot \mathbf{v}}{\mu} \mathbf{v},
\end{equation}
where $\mathbf{r}$ is the position vector, $\mathbf{v}$ is the velocity vector, $\mathbf{h}$ is the specific angular momentum vector, equal to $\mathbf{r} \times \mathbf{v}$, $\mu = G(M_\text{BH} + m_\text{S301}) \approx G M_\text{BH}$ is the standard gravitational parameter. The angular position of the pericentre can be represented as
\begin{equation}
  \theta = \arctan \left(\frac{e_y}{e_x}\right).
\end{equation}
In the calculation, we take the plane defined by the position $\mathbf{r}$ and velocity $\mathbf{v}$ of the BD at its first pericentre as the reference plane, which facilitates the subsequent computation of the precession angle using the eccentricity vector. We use $\theta_0, \theta_1, \theta_2, \theta_3$ to denote the pericentre angular positions of S301 in the absence of X-MRIs, and $\theta_0^\prime, \theta_1^\prime, \theta_2^\prime, \theta_3^\prime$ to denote those in the presence of X-MRIs. Thus the precession angles for each orbital period are $\omega_1 = \theta_1 - \theta_0$, $\omega_2 = \theta_2 - \theta_1$, $\omega_3 = \theta_3 - \theta_2$, and $\omega_1^\prime = \theta_1^\prime - \theta_0^\prime$, $\omega_2^\prime = \theta_2^\prime - \theta_1^\prime$, $\omega_3^\prime = \theta_3^\prime - \theta_2^\prime$. Consequently, the orbit-averaged precession residual is
\begin{equation}
  \Delta \omega = \frac{(\omega_1^\prime - \omega_1) + (\omega_2^\prime - \omega_2) + (\omega_3^\prime - \omega_3)}{3}.
\end{equation}
Dividing this by the orbital period of S301 yields the average precession deviation per year $\dot{\omega}$. {Because this residual is computed as the exact difference between the simulation with the X-MRIs and the baseline simulation without them, and since both simulations contain the exact same 37 S-stars, the secular gravitational influence of the S-stars cancels out perfectly. The residual is therefore attributed entirely to the X-MRIs.} Under the same condition of Fig.~\ref{fig:stellar3d}, we get one precession residual result $\dot{\omega} \approx 0.966\, \mathrm{arcsec\,yr}^{-1}$.

\begin{figure}[t]
  \centering
  \includegraphics[width=\columnwidth]{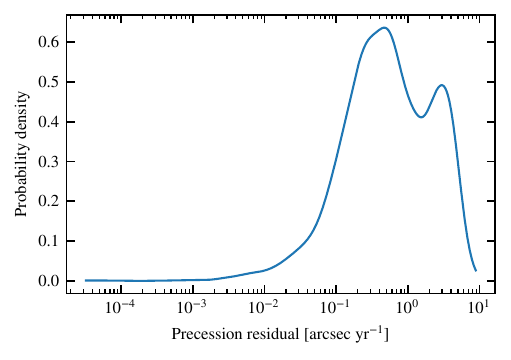}
  \caption{Distribution of the absolute precession residuals for S301 $\left|\dot{\omega}\right|$, which are predominantly concentrated between 0.1 and 10 $\mathrm{arcsec\,yr}^{-1}$.}
  \label{fig:distribution}
\end{figure}

Since the critical semi-major axis for X-MRIs $a_{\rm crit}$ is about $600\, \mathrm{AU}$, we uniformly sample 8 values within the range $[50\, \mathrm{AU}, 600\,\mathrm{AU}]$ for the semi-major axes of the two X-MRIs. Similarly, 8 values are uniformly sampled within the eccentricity range $[0.50, 0.98]$ {(while fixing the boundaries to the simulated {extrema} could marginally truncate distributions, future expansive continuous analyses can relax this interval without altering the qualitative results)}, resulting in a total of $8^4 = 4096$ parameter combinations. For each configuration, we perform numerical simulations following the procedure described above. The resulting precession residuals are then converted to absolute values and their distribution across different intervals is visualized using a kernel density estimation (KDE) plot (see Fig.~\ref{fig:distribution}).

After {averaging} over all these results, the resulting mean precession residual is
\begin{equation}
  \left\langle \dot{\omega} \right\rangle = \frac{1}{K}\sum^{K}_{k=1} \left|\dot{\omega}_k\right| = 1.15\, \mathrm{arcsec\,yr^{-1}},
  \label{eq:numericalprecession}
\end{equation}
where $K = 4096$. {The roughly 20\% deviation between the numerical simulations and the theoretical value ($0.963 \,\mathrm{arcsec\,yr}^{-1}$) is not caused by the S-stars, which are perfectly differenced out. Rather, the purely secular analytical approximation assumes an infinite-time continuous mass ring. In the N-body simulation, having just two discrete X-MRIs with periods comparable to S301 induces strong short-period fluctuations, specific phase dependencies, and higher-order multipole interactions that do not trivially average out over the relatively short 35-year integration limit.}

\subsection {Misalignment Case: A Qualitative Analysis of the S301 Precession Rate}

The above provides the precession rate of S301 induced by X-MRIs under the condition of perfect alignment. We proceed to calculate the results for the misaligned case. As an example, we choose inclination $\sigma_I = 45^\circ$, which is an intermediate, statistically representative value. Thus $\kappa \approx 3.24$, $\boldsymbol{\mathcal{R}}_I(\kappa) \approx 0.36$ calculated by Eq.~\eqref{eq:R_factor_exact}.

\begin{figure}[t]
  \centering
  \includegraphics[width=\columnwidth]{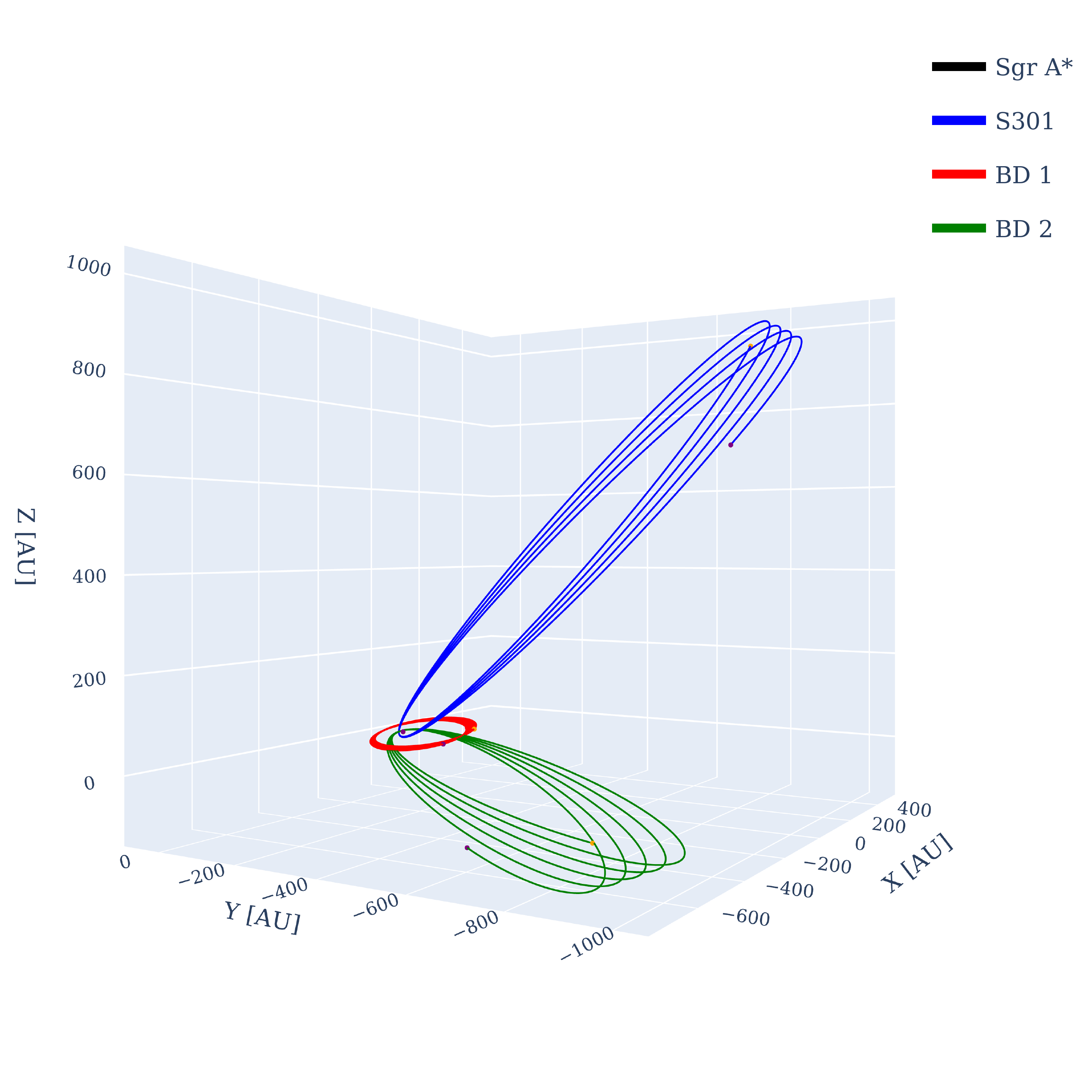}
  \caption{Evolutionary trajectories of the system comprising Sgr~A*, S301, and two BDs. Two X-MRIs are in the same plane, but this plane is inclined by $45^\circ$ relative to the orbital plane of S301.}
  \label{fig:stellar3d_45deg}
\end{figure}

In order to make it visual, we plot the orbital trajectories below, Fig.~\ref{fig:stellar3d_45deg}, where all parameters are identical to those in Fig.~\ref{fig:stellar3d} except for the inclination. {Specifically, the orientation of the orbital angular momentum vector $\mathbf{h} = \mathbf{r} \times \mathbf{v}$ and the $45^\circ$ rotation of the BDs' orbital plane relative to that of S301 follow the standard right-hand rule.}

\begin{figure}[t]
  \centering
  \includegraphics[width=\columnwidth]{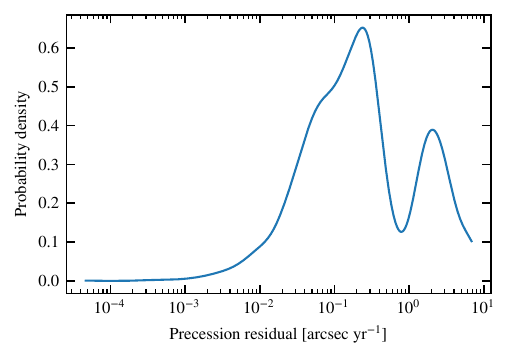}
  \caption{Distribution of the absolute precession residuals for S301
           $|\dot{\omega}|$ in the misalignment case, which are predominantly
           concentrated between 0.1 and 10 $\mathrm{arcsec\,yr}^{-1}$ as well.}
  \label{fig:distribution_45deg}
\end{figure}

{Following the same procedure as in} Fig.~\ref{fig:distribution} and Eq.~\eqref{eq:numericalprecession}, we {obtain the} KDE plot shown in Fig.~\ref{fig:distribution_45deg}, and the numerical mean precession residual for the misalignment case, $0.807 \, \mathrm{arcsec \, yr^{-1}}$, respectively. This simulated result is also in the same order of magnitude as the theoretical prediction, $0.344 \,\mathrm{arcsec\,yr}^{-1}$ calculated from Eq.~\eqref{eq:quad_precession_ext_disk}, demonstrating that the X-MRI signal is strong against misalignment, albeit with a reduced amplitude. {Similarly, the larger relative deviation ($>100\%$) from the theoretical value in this misaligned case arises because the purely secular quadrupole term is weakened, allowing the discrete, short-period N-body interactions to heavily dominate the dynamics. The results confirm that the method is robust for misaligned configurations, provided the S-stars are appropriately differenced or modeled.}

This establishes {an} approach based on {an} extensive exploration of the parameter space.

\section {Conclusions}

{This work establishes an approach for the detection of the hypothesized X-MRI population at the Galactic Centre using high-precision monitoring of S-star orbits. We provide an extensive analysis of the sensitivity of the recently detected star S301 \citep{S3012026} to both the interior X-MRIs and the exterior population of stellar-mass black holes (SBHs). We present a derivation showing that S301 is highly sensitive to the interior X-MRIs but largely insensitive to the secular effects of the exterior SBH population, caused by the rapid decay of the exterior perturbation with distance. While a single precession measurement is degenerate with respect to the exact number, mass, and inclination of the X-MRIs, measuring this torque proves the existence of the population and establishes strict bulk constraints on its mass prior to the launch of space-based GW detectors like LISA, TianQin, and Taiji.}

{The fundamental challenge stems from the granularity of the gravitational field caused by the SBHs. This population generates stochastic gravitational kicks that can mimic the slow, smooth secular precession caused by the X-MRIs. Standard analysis techniques that assume smoothness are biased by these kicks (\citealt{SadunBordoniEtAl2025}). We introduce a methodology centered on wavelet analysis, specifically the WTMM method, to overcome this ambiguity. By measuring the local smoothness of the orbital signal, we can distinguish between sharp stochastic kicks and smooth secular drifts, even in the presence of highly irregular measurement noise.}

{The practical implications of this study for data analysts are profound. The core message is that conventional orbital fitting, which seeks to minimize residuals against a smooth model, is insufficient in a granular environment. A statistically significant detection of precession using standard methods does not guarantee accuracy; it might simply be a biased measurement caused by a discrete kick. To detect X-MRIs, we must adopt a singularity-aware approach. I.e. (1) Analyze the texture of the residuals by measuring their local regularity using the CWT and the WTMM method. (2) Search for thrusts: a slope of $3/2$ (corresponding to $h=1$) is the unambiguous signature of a velocity discontinuity. (3) Validate smoothness assumptions: If a thrust is detected, the assumption of a globally smooth potential is violated. (4) Implement piecewise fitting: If a thrust is identified at time $t_k$, the analyst must {segment} the data. A piecewise model must be fitted, allowing for a discontinuity in the orbital elements at $t_k$. Only by explicitly identifying and removing the effects of granularity can the true secular precession be accurately measured.}

{The X-MRI signal is strong against misalignment: our $N$-body simulations yield a mean precession residual of $1.15 \, \mathrm{arcsec \, yr}^{-1}$ under perfect alignment, decreasing only to $0.807 \, \mathrm{arcsec \, yr}^{-1}$ at an inclination of $45^\circ$. These deviations from theoretical estimates are driven by short-period discrete N-body interactions rather than unmodeled S-stars, which are deterministically subtracted via our differential methodology. We perform a Fisher Matrix analysis, detailing how the precision scales with orbital parameters. We found that RV measurements dominate the error for S301 resulting from its high eccentricity. The detection timescale is rapid: a minimal population of two X-MRIs is detectable within $T_{4\sigma} \approx 8.7$ years. This wavelet-based validation is the bridge between the statistical precision offered by modern instruments and the astrophysical accuracy required to serve as a robust precursor for gravitational wave astronomy.}

\begin{acknowledgments}
PAS acknowledges support by China's National Foreign Expert Program (H). DC and JC acknowledge the financial support from ANID-FONDECYT Regular 1251444. The research of DC has been funded by the Alexander von Humboldt Foundation. QS acknowledges financial support from the China Scholarship Council (CSC) during his visit to València.
\end{acknowledgments}

\software{ARCHAIN \citep{1999CeMDA..74..287M, 1999MNRAS.310..745M, 2002CeMDA..84..343M, 2008AJ....135.2398M}}


\begin{thebibliography}{}
\expandafter\ifx\csname natexlab\endcsname\relax\def\natexlab#1{#1}\fi
\providecommand{\url}[1]{\href{#1}{#1}}
\providecommand{\dodoi}[1]{doi:~\href{http://doi.org/#1}{\nolinkurl{#1}}}
\providecommand{\doeprint}[1]{\href{http://ascl.net/#1}{\nolinkurl{http://ascl.net/#1}}}
\providecommand{\doarXiv}[1]{\href{https://arxiv.org/abs/#1}{\nolinkurl{https://arxiv.org/abs/#1}}}

\bibitem[{A. Akansu \& R. Haddad(2001)Akansu \&
  Haddad}]{akansu2001multiresolution}
Akansu, A., \& Haddad, R. 2001, Multiresolution Signal Decomposition:
  Transforms, Subbands, and Wavelets, Electronics \& Electrical (Elsevier
  Science)

\bibitem[{P. {Amaro-Seoane}(2019){Amaro-Seoane}}]{Amaro-Seoane2019}
{Amaro-Seoane}, P. 2019, \bibinfo{title}{{Extremely large mass-ratio
  inspirals},} Phys.Rev.D., 99, 123025, \dodoi{10.1103/PhysRevD.99.123025}

\bibitem[{P. {Amaro-Seoane} {et~al.}(2007){Amaro-Seoane}, {Gair}, {Freitag},
  {Miller}, {Mandel}, {Cutler}, \& {Babak}}]{Amaro-SeoaneEtAl07}
{Amaro-Seoane}, P., {Gair}, J.~R., {Freitag}, M., {et~al.} 2007,
  \bibinfo{title}{{Intermediate and extreme mass-ratio inspirals.
  Astrophysics,\ science applications and detection using LISA},} Classical and
  Quantum Gravity, 24, 113, \dodoi{10.1088/0264-9381/24/17/R01}

\bibitem[{P. {Amaro Seoane} {et~al.}(2024){Amaro Seoane}, {Lin}, \&
  {Tzanavaris}}]{AmaroSeoaneEtAl2024mono}
{Amaro Seoane}, P., {Lin}, Y., \& {Tzanavaris}, K. 2024, \bibinfo{title}{{Mono-
  and oligochromatic extreme-mass-ratio inspirals},} \prd, 110, 064011,
  \dodoi{10.1103/PhysRevD.110.064011}

\bibitem[{P. {Amaro Seoane} \& S.-D. {Zhao}(2025){Amaro Seoane} \&
  {Zhao}}]{AmaroSeoaneZhao2025}
{Amaro Seoane}, P., \& {Zhao}, S.-D. 2025, \bibinfo{title}{{A forest of
  gravitational waves in our Galactic Centre},} arXiv e-prints,
  arXiv:2504.20147, \dodoi{10.48550/arXiv.2504.20147}

\bibitem[{P. {Amaro-Seoane} {et~al.}(2017){Amaro-Seoane}, {Audley}, {Babak},
  {Baker}, {Barausse}, {Bender}, {Berti}, {Binetruy}, {Born}, {Bortoluzzi},
  {Camp}, {Caprini}, {Cardoso}, {Colpi}, {Conklin}, {Cornish}, {Cutler},
  {Danzmann}, {Dolesi}, {Ferraioli}, {Ferroni}, {Fitzsimons}, {Gair}, {Gesa
  Bote}, {Giardini}, {Gibert}, {Grimani}, {Halloin}, {Heinzel}, {Hertog},
  {Hewitson}, {Holley-Bockelmann}, {Hollington}, {Hueller}, {Inchauspe},
  {Jetzer}, {Karnesis}, {Killow}, {Klein}, {Klipstein}, {Korsakova}, {Larson},
  {Livas}, {Lloro}, {Man}, {Mance}, {Martino}, {Mateos}, {McKenzie},
  {McWilliams}, {Miller}, {Mueller}, {Nardini}, {Nelemans}, {Nofrarias},
  {Petiteau}, {Pivato}, {Plagnol}, {Porter}, {Reiche}, {Robertson},
  {Robertson}, {Rossi}, {Russano}, {Schutz}, {Sesana}, {Shoemaker}, {Slutsky},
  {Sopuerta}, {Sumner}, {Tamanini}, {Thorpe}, {Troebs}, {Vallisneri},
  {Vecchio}, {Vetrugno}, {Vitale}, {Volonteri}, {Wanner}, {Ward}, {Wass},
  {Weber}, {Ziemer}, \& {Zweifel}}]{Amaro-SeoaneEtAl2017}
{Amaro-Seoane}, P., {Audley}, H., {Babak}, S., {et~al.} 2017,
  \bibinfo{title}{{Laser Interferometer Space Antenna},} ArXiv e-prints.
\newblock \doarXiv{1702.00786}

\bibitem[{C. Chui(1992)Chui}]{chui1992introduction}
Chui, C. 1992, An Introduction to Wavelets, An Introduction to Wavelets No. v.
  1 (Elsevier Science)

\bibitem[{M. {Colpi} {et~al.}(2024){Colpi}, {Danzmann}, {Hewitson},
  {Holley-Bockelmann}, {Jetzer}, {Nelemans}, {Petiteau}, {Shoemaker},
  {Sopuerta}, {Stebbins}, {Tanvir}, {Ward}, {Weber}, {Thorpe}, {Daurskikh},
  {Deep}, {Fern{\'a}ndez N{\'u}{\~n}ez}, {Garc{\'\i}a Marirrodriga}, {Gehler},
  {Halain}, {Jennrich}, {Lammers}, {Larra{\~n}aga}, {Lieser},
  {L{\"u}tzgendorf}, {Martens}, {Mondin}, {Piris Ni{\~n}o}, {Amaro-Seoane},
  {Arca Sedda}, {Auclair}, {Babak}, {Baghi}, {Baibhav}, {Baker}, {Bayle},
  {Berry}, {Berti}, {Boileau}, {Bonetti}, {Brito}, {Buscicchio}, {Calcagni},
  {Capelo}, {Caprini}, {Caputo}, {Castelli}, {Chen}, {Chen}, {Chua}, {Davies},
  {Derdzinski}, {Domcke}, {Doneva}, {Dvorkin}, {Mar{\'\i}a Ezquiaga}, {Gair},
  {Haiman}, {Harry}, {Hartwig}, {Hees}, {Heffernan}, {Husa},
  {Izquierdo-Villalba}, {Karnesis}, {Klein}, {Korol}, {Korsakova}, {Kupfer},
  {Laghi}, {Lamberts}, {Larson}, {Le Jeune}, {Lewicki}, {Littenberg}, {Madge},
  {Mangiagli}, {Marsat}, {Vilchez}, {Maselli}, {Mathews}, {van de Meent},
  {Muratore}, {Nardini}, {Pani}, {Peloso}, {Pieroni}, {Pound},
  {Quelquejay-Leclere}, {Ricciardone}, {Rossi}, {Sartirana}, {Savalle},
  {Sberna}, {Sesana}, {Shoemaker}, {Slutsky}, {Sotiriou}, {Speri}, {Staab},
  {Steer}, {Tamanini}, {Tasinato}, {Torrado}, {Torres-Orjuela}, {Toubiana},
  {Vallisneri}, {Vecchio}, {Volonteri}, {Yagi}, \& {Zwick}}]{ColpiEtAl2024}
{Colpi}, M., {Danzmann}, K., {Hewitson}, M., {et~al.} 2024,
  \bibinfo{title}{{LISA Definition Study Report},} arXiv e-prints,
  arXiv:2402.07571, \dodoi{10.48550/arXiv.2402.07571}

\bibitem[{B. Cordani(2003)Cordani}]{Cordani2003}
Cordani, B. 2003, The Kepler Problem (Birkh\"{a}user Basel),
  \dodoi{10.1007/978-3-0348-8051-0}

\bibitem[{I. Daubechies(1992)Daubechies}]{daubechies1992ten}
Daubechies, I. 1992, Ten Lectures on Wavelets, CBMS-NSF Regional Conference
  Series in Applied Mathematics (Society for Industrial and Applied
  Mathematics)

\bibitem[{K.~A. {El Dayem} {et~al.}(2026){El Dayem}, {Abuter}, {Aimar},
  {Amaro-Seoane}, {Berdeu}, {Berger}, {Bourdarot}, {Brandner}, {Burkert},
  {Calderon}, {Correia}, {Cuadra}, {Davies}, {Defrere}, {Delit}, {Drescher},
  {Eisenhauer}, {Esteras Otal}, {Fabricius}, {Feuchtgruber}, {Foerster
  Schreiber}, {Foschi}, {Garcia}, {Garcia Lopez}, {Generozov}, {Genzel},
  {Gillessen}, {Gonte}, {Haubois}, {Hoenig}, {Houlle}, {Joharle}, {Kaufer},
  {Kammerer}, {Kervella}, {Kolb}, {Kreidberg}, {Labadie}, {Lacour}, {Lai},
  {Laugier}, {Le Bouquin}, {Leftley}, {Lopez}, {Lutz}, {Mang}, {Merand},
  {Millour}, {Montarges}, {Morujao}, {Nowacki}, {Nowak}, {Oberti}, {Osorno},
  {Ott}, {Paumard}, {Paladini}, {Perets}, {Perraut}, {Perrin}, {Petrov},
  {Petrucci}, {Piran}, {Pourre}, {Rabien}, {Ribeiro}, {Robbe-Dubois}, {Sadun
  Bordoni}, {Sanchez Bermudez}, {Santos}, {Sari}, {Sauter}, {Scheithauer},
  {Scigliuto}, {Shangguan}, {Shimizu}, {Soulez}, {Stadler}, {Straubmeier},
  {Sturm}, {Subroweit}, {Sykes}, {Tacconi}, {Thevenet}, {Urso}, {Vincent},
  {Woillez}, \& {Zins}}]{S3012026}
{El Dayem}, K.~A., {Abuter}, R., {Aimar}, N., {et~al.} 2026,
  \bibinfo{title}{{Discovery of a star sensitive to the spin of Sgr A*},} arXiv
  e-prints, arXiv:2607.12664.
\newblock \doarXiv{2607.12664}

\bibitem[{H.-M. {Fan} {et~al.}(2026){Fan}, {Torres-Orjuela},
  {V{\'a}zquez-Aceves}, {Wang}, \& {Feng}}]{FanEtAl2026}
{Fan}, H.-M., {Torres-Orjuela}, A., {V{\'a}zquez-Aceves}, V., {Wang}, T.-X., \&
  {Feng}, T.-F. 2026, \bibinfo{title}{{Constraining modified theories of
  gravity through the detection of one extremely large mass-ratio inspiral},}
  \prd, 113, 104021, \dodoi{10.1103/slrf-wl5c}

\bibitem[{R.~A. {Fisher}(1920){Fisher}}]{Fisher1920}
{Fisher}, R.~A. 1920, \bibinfo{title}{{Accuracy of observation, A mathematical
  examination of the methods of determining, by the mean error and by the mean
  square error},} \mnras, 80, 758, \dodoi{10.1093/mnras/80.8.758}

\bibitem[{R.~A. {Fisher}(1922){Fisher}}]{Fisher1922}
{Fisher}, R.~A. 1922, \bibinfo{title}{{On the Mathematical Foundations of
  Theoretical Statistics},} Philosophical Transactions of the Royal Society of
  London Series A, 222, 309, \dodoi{10.1098/rsta.1922.0009}

\bibitem[{R.~A. {Fisher}(1925){Fisher}}]{Fisher1925}
{Fisher}, R.~A. 1925, \bibinfo{title}{{Theory of Statistical Estimation},}
  Proceedings of the Cambridge Philosophical Society, 22, 700,
  \dodoi{10.1017/S0305004100009580}

\bibitem[{ {GRAVITY Collaboration} {et~al.}(2020){GRAVITY Collaboration},
  {Abuter}, {Amorim}, {Baub{\"o}ck}, {Berger}, {Bonnet}, {Brandner}, {Cardoso},
  {Cl{\'e}net}, {de Zeeuw}, {Dexter}, {Eckart}, {Eisenhauer}, {F{\"o}rster
  Schreiber}, {Garcia}, {Gao}, {Gendron}, {Genzel}, {Gillessen}, {Habibi},
  {Haubois}, {Henning}, {Hippler}, {Horrobin}, {Jim{\'e}nez-Rosales}, {Jochum},
  {Jocou}, {Kaufer}, {Kervella}, {Lacour}, {Lapeyr{\`e}re}, {Le Bouquin},
  {L{\'e}na}, {Nowak}, {Ott}, {Paumard}, {Perraut}, {Perrin}, {Pfuhl},
  {Rodr{\'\i}guez-Coira}, {Shangguan}, {Scheithauer}, {Stadler}, {Straub},
  {Straubmeier}, {Sturm}, {Tacconi}, {Vincent}, {von Fellenberg}, {Waisberg},
  {Widmann}, {Wieprecht}, {Wiezorrek}, {Woillez}, {Yazici}, \&
  {Zins}}]{GRAVITY2020}
{GRAVITY Collaboration}, {Abuter}, R., {Amorim}, A., {et~al.} 2020,
  \bibinfo{title}{{Detection of the Schwarzschild precession in the orbit of
  the star S2 near the Galactic centre massive black hole},} \aap, 636, L5,
  \dodoi{10.1051/0004-6361/202037813}

\bibitem[{ {Gravity Collaboration} {et~al.}(2024){Gravity Collaboration}, {Abd
  El Dayem}, {Abuter}, {Aimar}, {Amaro Seoane}, {Amorim}, {Beck}, {Berger},
  {Bonnet}, {Bourdarot}, {Brandner}, {Cardoso}, {Capuzzo Dolcetta},
  {Cl{\'e}net}, {Davies}, {de Zeeuw}, {Drescher}, {Eckart}, {Eisenhauer},
  {Feuchtgruber}, {Finger}, {F{\"o}rster Schreiber}, {Foschi}, {Gao}, {Garcia},
  {Gendron}, {Genzel}, {Gillessen}, {Hartl}, {Haubois}, {Haussmann},
  {Hei{\ss}el}, {Henning}, {Hippler}, {Horrobin}, {Jochum}, {Jocou}, {Kaufer},
  {Kervella}, {Lacour}, {Lapeyr{\`e}re}, {Le Bouquin}, {L{\'e}na}, {Lutz},
  {Mang}, {More}, {Ott}, {Paumard}, {Perraut}, {Perrin}, {Pfuhl}, {Rabien},
  {Ribeiro}, {Sadun Bordoni}, {Scheithauer}, {Shangguan}, {Shimizu}, {Stadler},
  {Straub}, {Straubmeier}, {Sturm}, {Tacconi}, {Urso}, {Vincent}, {von
  Fellenberg}, {Widmann}, {Wieprecht}, {Woillez}, \& {Zhang}}]{GRAVITY2024}
{Gravity Collaboration}, {Abd El Dayem}, K., {Abuter}, R., {et~al.} 2024,
  \bibinfo{title}{{Improving constraints on the extended mass distribution in
  the Galactic center with stellar orbits},} \aap, 692, A242,
  \dodoi{10.1051/0004-6361/202452274}

\bibitem[{E.-K. {Li} {et~al.}(2025){Li}, {Liu}, {Torres-Orjuela}, {Chen},
  {Inayoshi}, {Wang}, {Hu}, {Amaro-Seoane}, {Askar}, {Bambi}, {Capelo}, {Chen},
  {Chua}, {Cond{\'e}s-Bre{\~n}a}, {Dai}, {Das}, {Derdzinski}, {Fan}, {Fujii},
  {Gao}, {Garg}, {Ge}, {Giersz}, {Huang}, {Hypki}, {Liang}, {Liu}, {Liu},
  {Liu}, {Liu}, {Mayer}, {Napolitano}, {Peng}, {Shao}, {Shashank}, {Shen},
  {Tagawa}, {Tanikawa}, {Toscani}, {V{\'a}zquez-Aceves}, {Wang}, {Wang}, {Yi},
  {Zhang}, {Zhang}, {Zhu}, {Zwick}, {Huang}, {Mei}, {Wang}, {Xie}, {Zhang}, \&
  {Luo}}]{TianQin2025}
{Li}, E.-K., {Liu}, S., {Torres-Orjuela}, A., {et~al.} 2025,
  \bibinfo{title}{{Gravitational wave astronomy with TianQin},} Reports on
  Progress in Physics, 88, 056901, \dodoi{10.1088/1361-6633/adc9be}

\bibitem[{J. {Luo} {et~al.}(2016){Luo}, {Chen}, {Duan}, {Gong}, {Hu}, {Ji},
  {Liu}, {Mei}, {Milyukov}, {Sazhin}, {Shao}, {Toth}, {Tu}, {Wang}, {Wang},
  {Yeh}, {Zhan}, {Zhang}, {Zharov}, \& {Zhou}}]{LuoEtAl2016}
{Luo}, J., {Chen}, L.-S., {Duan}, H.-Z., {et~al.} 2016,
  \bibinfo{title}{{TianQin: a space-borne gravitational wave detector},}
  Classical and Quantum Gravity, 33, 035010,
  \dodoi{10.1088/0264-9381/33/3/035010}

\bibitem[{S. Mallat(1999)Mallat}]{mallat1999wavelet}
Mallat, S. 1999, A Wavelet Tour of Signal Processing, Wavelet Analysis and Its
  Applications (Academic Press)

\bibitem[{S. {Mallat} \& W.~L. {Hwang}(1992){Mallat} \&
  {Hwang}}]{MallatHwang1992}
{Mallat}, S., \& {Hwang}, W.~L. 1992, \bibinfo{title}{{Singularity detection
  and processing with wavelets},} IEEE Transactions on Information Theory, 38,
  617, \dodoi{10.1109/18.119727}

\bibitem[{Y. Meyer(1993)Meyer}]{Meyer_1993}
Meyer, Y. 1993, Wavelets and Operators, ed. D.~H. Salinger, Cambridge Studies
  in Advanced Mathematics (Cambridge University Press)

\bibitem[{S. {Mikkola} \& S. {Aarseth}(2002){Mikkola} \&
  {Aarseth}}]{2002CeMDA..84..343M}
{Mikkola}, S., \& {Aarseth}, S. 2002, \bibinfo{title}{{A Time-Transformed
  Leapfrog Scheme},} Celestial Mechanics and Dynamical Astronomy, 84, 343,
  \dodoi{10.1023/A:1021149313347}

\bibitem[{S. {Mikkola} \& D. {Merritt}(2008){Mikkola} \&
  {Merritt}}]{2008AJ....135.2398M}
{Mikkola}, S., \& {Merritt}, D. 2008, \bibinfo{title}{{Implementing Few-Body
  Algorithmic Regularization with Post-Newtonian Terms},} \aj, 135, 2398,
  \dodoi{10.1088/0004-6256/135/6/2398}

\bibitem[{S. {Mikkola} \& K. {Tanikawa}(1999{\natexlab{a}}){Mikkola} \&
  {Tanikawa}}]{1999CeMDA..74..287M}
{Mikkola}, S., \& {Tanikawa}, K. 1999{\natexlab{a}}, \bibinfo{title}{{Explicit
  Symplectic Algorithms For Time-Transformed Hamiltonians},} Celestial
  Mechanics and Dynamical Astronomy, 74, 287, \dodoi{10.1023/A:1008368322547}

\bibitem[{S. {Mikkola} \& K. {Tanikawa}(1999{\natexlab{b}}){Mikkola} \&
  {Tanikawa}}]{1999MNRAS.310..745M}
{Mikkola}, S., \& {Tanikawa}, K. 1999{\natexlab{b}},
  \bibinfo{title}{{Algorithmic regularization of the few-body problem},}
  \mnras, 310, 745, \dodoi{10.1046/j.1365-8711.1999.02982.x}

\bibitem[{C.~D. {Murray} \& S.~F. {Dermott}(1999){Murray} \&
  {Dermott}}]{MurrayDermott1999}
{Murray}, C.~D., \& {Dermott}, S.~F. 1999, {Solar System Dynamics} ({Cambridge
  University Press}), \dodoi{10.1017/CBO9781139174817}

\bibitem[{K.~P. {Rauch} \& S. {Tremaine}(1996){Rauch} \& {Tremaine}}]{RT96}
{Rauch}, K.~P., \& {Tremaine}, S. 1996, \bibinfo{title}{Resonant relaxation in
  stellar systems,} New Astronomy, 1, 149

\bibitem[{W.-H. {Ruan} {et~al.}(2020){Ruan}, {Guo}, {Cai}, \&
  {Zhang}}]{Taiji2020}
{Ruan}, W.-H., {Guo}, Z.-K., {Cai}, R.-G., \& {Zhang}, Y.-Z. 2020,
  \bibinfo{title}{{Taiji program: Gravitational-wave sources},} International
  Journal of Modern Physics A, 35, 2050075, \dodoi{10.1142/S0217751X2050075X}

\bibitem[{M. {Sadun Bordoni} {et~al.}(2025){Sadun Bordoni}, {Capuzzo Dolcetta},
  {Generozov}, {Bourdarot}, {Drescher}, {Eisenhauer}, {Genzel}, {Gillessen},
  {Joharle}, {Mang}, {Ott}, {Ribeiro}, \& {von
  Fellenberg}}]{SadunBordoniEtAl2025}
{Sadun Bordoni}, M., {Capuzzo Dolcetta}, R., {Generozov}, A., {et~al.} 2025,
  \bibinfo{title}{{Impact of a granular mass distribution on the orbit of S2 in
  the Galactic center},} \aap, 701, A89, \dodoi{10.1051/0004-6361/202554223}

\bibitem[{B. Taylor(1715)Taylor}]{Taylor1715}
Taylor, B. 1715, Methodus Incrementorum Directa et Inversa (Londini : Typis
  Pearsonianis prostant apud Gul. Innys ad Insignia Principis in Coemeterio
  Paulino).
\newblock \url{https://archive.org/details/UFIE003454_TO0324_PNI-2529_000000}

\bibitem[{V. {V{\'a}zquez-Aceves} {et~al.}(2023){V{\'a}zquez-Aceves}, {Lin}, \&
  {Torres-Orjuela}}]{VazquezEtAl2023}
{V{\'a}zquez-Aceves}, V., {Lin}, Y., \& {Torres-Orjuela}, A. 2023,
  \bibinfo{title}{{Sgr A* Spin and Mass Estimates through the Detection of an
  Extremely Large Mass-ratio Inspiral},} \apj, 952, 139,
  \dodoi{10.3847/1538-4357/acde51}

\bibitem[{V. {V{\'a}zquez-Aceves} {et~al.}(2025){V{\'a}zquez-Aceves}, {Lin}, \&
  {Torres-Orjuela}}]{VazquezEtAl2025}
{V{\'a}zquez-Aceves}, V., {Lin}, Y., \& {Torres-Orjuela}, A. 2025,
  \bibinfo{title}{{Sgr A* Spin and Mass Estimates through the Detection of
  Multiple Extremely Large Mass Ratio Inspirals},} \apj, 987, 208,
  \dodoi{10.3847/1538-4357/ade149}

\bibitem[{V. {V{\'a}zquez-Aceves} {et~al.}(2022){V{\'a}zquez-Aceves}, {Zwick},
  {Bortolas}, {Capelo}, {Amaro Seoane}, {Mayer}, \& {Chen}}]{VazquezEtAl2022}
{V{\'a}zquez-Aceves}, V., {Zwick}, L., {Bortolas}, E., {et~al.} 2022,
  \bibinfo{title}{{Revised event rates for extreme and extremely large
  mass-ratio inspirals},} \mnras, 510, 2379, \dodoi{10.1093/mnras/stab3485}

\bibitem[{T.-X. {Wang} {et~al.}(2026){Wang}, {Wang}, {Torres-Orjuela}, {Lin},
  {Fan}, {V{\'a}zquez-Aceves}, \& {Hu}}]{WangEtAl2026}
{Wang}, T.-X., {Wang}, Y., {Torres-Orjuela}, A., {et~al.} 2026,
  \bibinfo{title}{{Constructing a gravitational wave analysis pipeline for
  extremely large-mass-ratio inspirals},} \prd, 114, 023005,
  \dodoi{10.1103/3hx6-7zyb}

\end{thebibliography}
\end{document}